\documentclass[11pt,a4paper]{emulateapj}
\usepackage{courier,graphicx,epsfig,color,colordvi}

\usepackage{textcomp}
\usepackage{txfonts}
\usepackage{time}
\usepackage{natbib,twoopt}
\usepackage[breaklinks=true,colorlinks=true,urlcolor=blue,citecolor=blue,linkcolor=blue]{hyperref} %% to avoid \citeads line fills

\bibpunct{(}{)}{;}{a}{}{,} %% natbib format for A&A and ApJ
\makeatletter
\newcommandtwoopt{\citeads}[3][][]{\href{http://adsabs.harvard.edu/abs/#3}%
{\def\hyper@linkstart##1##2{}%
\let\hyper@linkend\@empty\citealp[#1][#2]{#3}}}
\newcommandtwoopt{\citepads}[3][][]{\href{http://adsabs.harvard.edu/abs/#3}%
{\def\hyper@linkstart##1##2{}%
\let\hyper@linkend\@empty\citep[#1][#2]{#3}}}
\newcommandtwoopt{\citetads}[3][][]{\href{http://adsabs.harvard.edu/abs/#3}%
{\def\hyper@linkstart##1##2{}%
\let\hyper@linkend\@empty\citet[#1][#2]{#3}}}
\newcommandtwoopt{\citeyearads}[3][][]%
{\href{http://adsabs.harvard.edu/abs/#3}
{\def\hyper@linkstart##1##2{}%
\let\hyper@linkend\@empty\citeyear[#1][#2]{#3}}}
\makeatother

\definecolor{dark-gray}{gray}{0.4}
\begin{document}
%\authorrunning{de la Cruz Rodr\'igue et al.}

\title{Emergence of granular-sized magnetic bubbles through the solar
  atmosphere\\II. Non-LTE chromospheric diagnostics and inversions}
%\subtitle{}

\author{Jaime de la Cruz Rodr\'iguez\altaffilmark{1}}\email{jaime@astro.su.se}
\author{Viggo Hansteen\altaffilmark{2}}
\author{Luis Bellot-Rubio\altaffilmark{3}}
\author{Ada Ortiz\altaffilmark{2}}

\affil{\altaffilmark{1}Institute for Solar Physics, Dept. of Astronomy,
  Stockholm University, Albanova University Center,
  SE-10691 Stockholm, Sweden}
  \affil{\altaffilmark{2} Institute of
  Theoretical Astrophysics, University of Oslo, P.O. Box 1029
  Blindern, N--0315 Oslo, Norway}
  \affil{\altaffilmark{3} Instituto de Astrof\'{\i}sica de Andaluc\'{\i}a (CSIC), %
  Apartado de Correos 3004, ES-18080 Granada, Spain}

% \date{Received September 15, 1996; Accepted March 16, 1997}
%\date{Draft: \now\ \today}
\frenchspacing
\begin{abstract}
Magnetic flux emergence into the outer layers of the Sun is a
  fundamental mechanism for releasing energy into the chromosphere and
  the corona. In this paper, we study the emergence of
  granular-sized flux concentrations and the structuring of the corresponding
  physical parameters and atmospheric diagnostics in the upper
  photosphere and in the chromosphere. We make use of a realistic 3D MHD simulation of the outer layers of
  the Sun to study the formation of the \ion{Ca}{2}~$8542$ line. We also derive semi-empirical 3D models from
  non-LTE inversions of our observations. These models contain
  depth-dependent information of the temperature and line-of-sight stratification.
Our analysis explains the peculiar \ion{Ca}{2}~$8542$~\AA\
  profiles observed in the flux-emerging region. In addition,
  we derive detailed temperature and velocity maps describing the ascent
  of magnetic bubbles from the photosphere to the chromosphere. The inversions suggest that, in active regions, granular-sized
  bubbles emerge up to the lower-chromosphere where the existing
  large-scale field hinders their ascent. We report hints of
  heating when the field reaches the
  chromosphere.
\end{abstract}

%\maketitle

\section{Introduction}\label{sec:intro}
Currently, one of the foremost questions in
solar physics concerns the mechanisms
that transport and release non-thermal energy into the outermost layers of
the Sun (e.g. \citeads{1992str..book.....M,2006SoPh..234...41K}). 
% Mariska's book and Klimchuks review... Enough?
 The answer to these questions is evidently connected
with understanding the fine structuring
observed in the chromosphere, because the conversion of non-thermal
energy into thermal energy occurs at very small spatial
scales (e.g.,\citeads{2014Sci...346A.315T};
\citeads{2014Sci...346E.315H}; \citeads{2014Sci...346D.315D}). Magnetic fields and magnetic
reconnection play important roles in heating the
chromosphere and corona (e.g. \citeads{1996JGR...10113445G,2005ApJ...618.1020G}).
Reconnection can occur as a result of the
braiding of the previously existing magnetic field by photospheric motions, but can be
strongly enhanced when new magnetic flux emerges through the
photosphere (\citeads{2007ApJ...666..516G}).
% Galsgaard et al 2007
The emergence of magnetic flux into the outer layers of the
Sun also plays an important role in the formation of sunspots and active regions,
and is an integral part of the solar cycle
(\citeads{2014SSRv..186..227S}).
% Schmieder et al 2014.

During the past years, a number of observational studies have aimed to describe how
small-scale magnetic fields emerge in the outer layers of the Sun, and to establish
observational constraints on the shape, field-strength and speed of ascent through the photosphere and lower chromosphere
(\citeads{2009ApJ...700.1391M}; \citeads{2014SSRv..186..227S}). In the
photosphere, great attention has been given to magnetic-flux
emergence in the quiet-Sun, because the ubiquity of
the latter would translate into large amounts of magnetic flux
being deposited into the outer layers of the atmosphere
(see recent studies by \citeads{2008ApJ...672.1237L};
\citeads{2009A&A...502..969B}; \citeads{2010ApJ...723L.149D};
\citeads{2014ApJ...797...49G}, and references therein).

In regions with relatively high magnetic activity, and therefore
stronger polarization signatures, observers have focussed on the
coupling between the photosphere and the lower chromosphere and have studied
the connectivity of the emerging flux (\citeads{2010ApJ...724.1083G};
\citeads{2014ApJ...781..126O}). However, not much work has been done
using chromospheric diagnostics, probably because
non-LTE/non-equilibrium conditions make it very difficult
to translate the observed intensities into the underlying physical
state of the atmosphere (see, e.g, \citeads{2000ApJ...530..977S};
\citeads{2012ApJ...749..136L}). 

Realistic radiative numerical MHD simulations of flux emergence have, in most cases, been
restricted below the upper photosphere
(e.g., \citeads{2008ApJ...687.1373C}; \citeads{2009A&A...507..949T})
because non-LTE radiative losses must be included to reproduce
chromospheric conditions. Using the non-LTE method developed by 
\citetads{2000ApJ...536..465S}, \citetads{2008ApJ...679..871M, 2009ApJ...702..129M}, studied for the first time the emergence of magnetic flux in a simulation
including a photosphere, chromosphere and corona. They investigated the
chromospheric and coronal response to the emergence of magnetic
flux, reporting the formation of cold magnetic bubbles in the upper
photosphere that expel chromospheric oscillations and pushes the
transition region and the corona to heights much greater than 2~Mm. 

A similar scenario is described by
\citetads{2014ApJ...781..126O} (hereafter Paper I), who compare \ion{Ca}{2}~IR
observations with the atmospheric parameters of a numerical
simulation. They find observational evidences of these cold bubbles,
which had been previously found in simulations by
\citetads{2008ApJ...679..871M, 2009ApJ...702..129M}. 
The ascent of a magnetic bubble from the photosphere into the 
chromosphere was described using very high spatial-resolution data in the 
\ion{Fe}{1}~$\lambda 6301/\lambda 6302$ and the \ion{Ca}{2}~$\lambda 8542$ lines. 
Furthermore, in Paper I the authors also presented photospheric LTE inversions from the \ion{Fe}{1} dataset,
and a qualitative study of the \ion{Ca}{2} data, including a weak-field approximation analysis to infer the 
strength of the magnetic field in the chromosphere.

In this paper, we investigate the emergence of small-scale magnetic flux
from the photosphere into the chromosphere, continuing the study initiated by
\citetads{2014ApJ...781..126O}. Our analysis is
prompted by our desire to know the structuring of
physical parameters in the upper photosphere and lower/middle
chromosphere during the emergence of a cold magnetic bubble.  We also wish to understand the peculiar shape of the
\ion{Ca}{2}~$\lambda8542$ intensity profiles observed in 
the flux emergence events, with a bump in the red wing that looks
like an emission peak. The former are obtained by performing non-LTE
inversions of high spatial-resolution observations, whereas the latter is investigated using a realistic 3D MHD simulation of the outer layers
of the sun.

In Sect.~\ref{sec:dat} we describe the processing of the observations, the inversions, and the computation of synthetic profiles from the 3D MHD simulations. In
Sect.~\ref{sec:resu} we analyze the inverted models, and we discuss the formation of the
\ion{Ca}{2}~$8542$ line during the flux emergence, using the 3D simulation. 
Our conclusions are summarized in Sect.~\ref{sec:conclu}.

\section{Data}\label{sec:dat}
\subsection{Observations with the SST}\label{sec:obs}
We analyze the same dataset that was used in Paper I. Our observations
of AR11024 were acquired on 2009-07-05 starting at 09:48 UT, at
coordinates S27$^\circ$, W12$^\circ$. The data were taken with the Swedish 1-m Solar Telescope
(SST, \citeads{2013A&A...553A..63S}) and the CRisp Imaging
Spectro-Polarimeter
(CRISP, \citeads{2008ApJ...689L..69S}) in the
\ion{Ca}{2}~$\lambda8542$~\AA\ line (hereafter $\lambda 8542$).

We sequentially acquired narrow band images sampling the
$\lambda8542$ profile at 17 line positions in steps of 100~m\AA \ between
$\Delta\lambda=\pm800$~m\AA \ from line center. An additional point
was measured at $\Delta \lambda=2400$~m\AA.

The data were reduced using the CRISPRED pipeline (\citeads{2015A&A...573A..40D}). The images
have been restored with multi-frame-multi-object-blind-deconvolution
techniques (MOMFBD, \citeads{2005SoPh..228..191V}) and residual seeing
distortions within the line scan have been compensated using methods
described in \citetads{2012A&A...548A.114H}.

It is not possible to reach the continuum of the $\lambda 8542$ line
with CRISP because the prefilter is not wide enough. Therefore, to
scale our observations relative to the continuum of the quiet-Sun spatial average at
disk-center, we use a 3D HD simulation
(\citeads{2011A&A...528A.113D}). The idea is to compute a simulated
spatio-temporal average of quiet-Sun spectra at $\mu=0.87$
(where our observations were acquired). Then we normalize the average
intensity profile by the continuum intensity at $\mu=1$. Finally, we
scaled our observations in such a way that the average quiet-Sun
spectrum at $\Delta\lambda = 2.4$~\AA\ has the same intensity as the normalised synthetic average profile. The same technique was used in Paper I and by
\citetads{2012ApJ...757...49W}.

\begin{figure}[]
  \centering
  \includegraphics[width=\hsize]{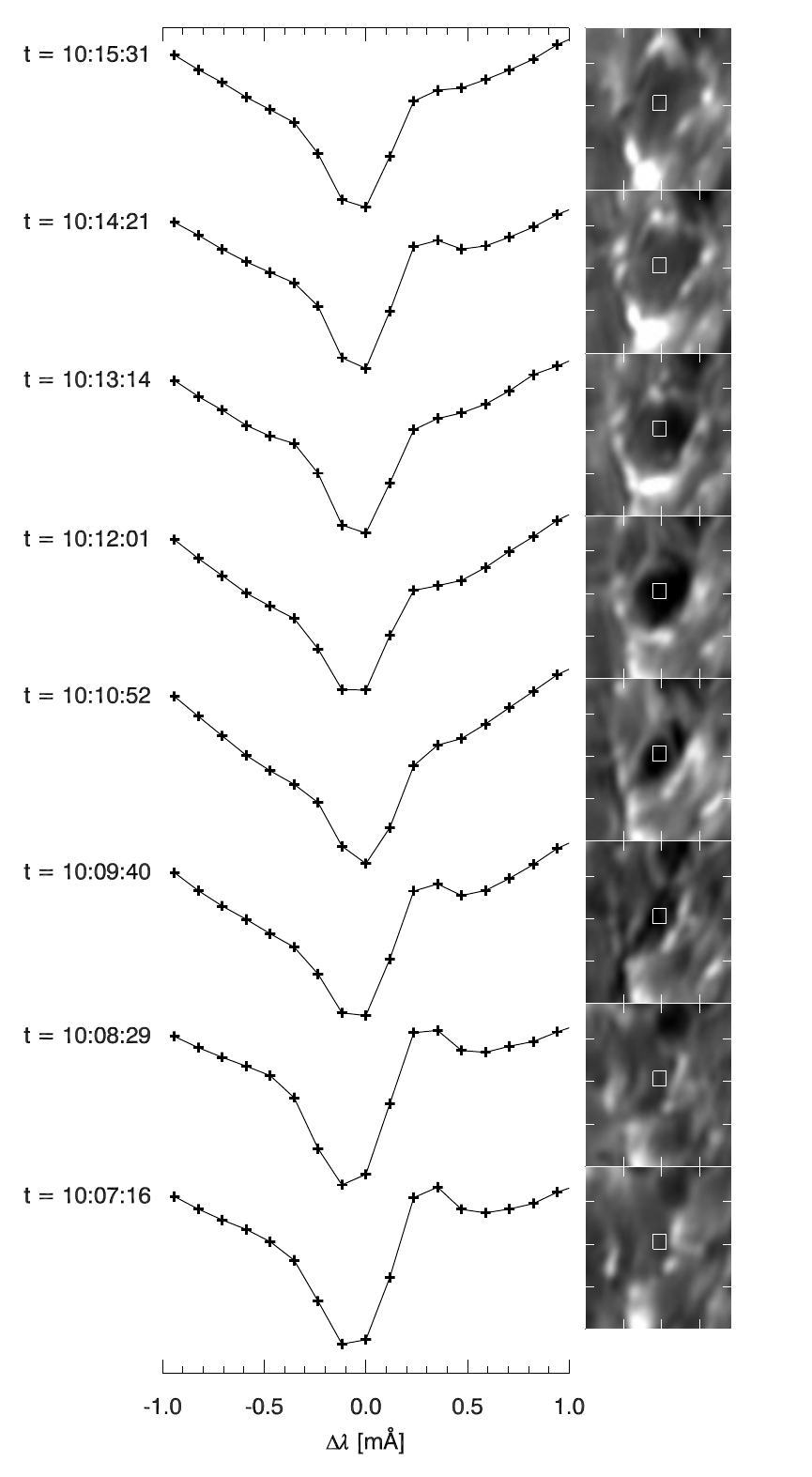}
  \caption{From bottom to top, time evolution of the $\lambda 8542$
    spectra, spatially averaged within the small white box ($7\times
    7$~pixels) indicated in the filtergrams on the right. The tick-mark
    separation is $1$\arcsec \ in the images.}
  \label{fig:time}
\end{figure}

In Paper I we described the peculiar intensity profile of the $\lambda8542$ line during
the flux emergence event. Figure~\ref{fig:time} illustrates a
time-series of the profile within the region of interest, indicated in
the corresponding images with a white square. The profile has three
distinctive features compared with a spatial quiet-Sun average:
\begin{enumerate}
  \item The line center is clearly blue-shifted.
  \item Close to line center, the line profile is strongly asymmetric, with a
    very sharp gradient on the red side, and a much more extended blue
    side.
  \item At approximately $+250$~m\AA, there is an
    emission feature that is not similarly visible on the blue wing.
\end{enumerate}
The degree of asymmetry and the exact strength of the three features
enumerated above suffer slight changes over time, but the overall
shape of the profile is persistently similar. In our analysis, we will try to
understand the formation of the $\lambda8542$
line and the atmospheric parameters that are responsible for these features.

\subsection{non-LTE calculations}\label{sec:nic}
We use the non-LTE inversion code
\textsc{Nicole} to compute inversions in the
$\lambda 8542$ line (\citeads{2014arXiv1408.6101S}). The code
iteratively modifies the
physical parameters of a \emph{guessed} model atmosphere, to reproduce
observed spectra (\citeads{2000ApJ...530..977S}). \textsc{Nicole}
solves the non-LTE problem as described in
\citetads{1997ApJ...490..383S} assuming statistical equilibrium, which
is reasonable for \ion{Ca}{2} lines (\citeads{2011A&A...528A...1W}).
Our atom model
consists of five bound levels plus
an ionization continuum, similar to that used by
\citetads{2009ApJ...694L.128L}. The atomic data used in this study has
been extracted from the VALD database
(\citeads{1995A&AS..112..525P}). 

\textsc{Nicole} assumes Zeeman-induced
polarization to compute the full-Stokes vector, neglecting scattering
polarization and the Hanle effect. This assumption works fine in
regions with relatively strong magnetic-field, but in the quiet-Sun
all these effects must be considered to model Stokes~$Q$ and $U$
observations (\citeads{2010ApJ...722.1416M};
\citeads{2012ApJ...751....5C}; \citeads{2014arXiv1412.5386C}).

The radiative transfer equation is solved using a piece-wise cubic
Bezier spline interpolant both to calculate the non-LTE populations
(assuming unpolarized light) and to compute the final polarized formal
solution (\citeads{2003ASPC..288....3A}; \citeads{2013ApJ...764...33D}).

The spatially-averaged quiet-Sun spectrum of the $\lambda 8542$ line
shows a reversed C-shape bisector
(\citeads{2006ApJ...639..516U}; \citeads{2013ApJ...764..153P}; \citeads{2013SoPh..288...89C}). Only recently,
\citetads{2014ApJ...784L..17L} realized that the asymmetry in solar
observations is produced
by the presence of different \ion{Ca}{2} isotopes, apparently a well-known phenomenon
in chemically peculiar stars (\citeads{2004A&A...421L...1C};
\citeads{2005A&A...432L..21C}; \citeads{2007MNRAS.377.1579C};
\citeads{2008A&A...480..811R}).

We have included the effect of isotopic splitting in our inversions
using a simple approximation. First, the non-LTE problem is solved with a
regular $^{40}$Ca atom. Once the atom population densities are known,
we compute a final formal solution of the radiative transfer equation, 
assuming that the absorption coefficient is a weighted sum of Voigt
profiles, each of them centered at the rest-wavelength of the line for
each isotope. In our case, the weight is proportional to the relative abundance of each isotope
(\citeads{1986UppOR..33.....C}).

Our observations are under-sampled by almost a factor $\times 2$ in the
spectral dimension, a necessary trade-off to keep the cadence
sufficiently high while observing three spectral lines. This situation
is less than ideal for inversions, where critically sampled
observations constrain the solution much better. Therefore we have
run a set of tests to assess the optimal combination of nodes that
allows us to reproduce the observed profiles with the lowest number of
degrees of freedom that is possible. The inversions are performed in
several cycles to improve the convergence (\citeads{1992ApJ...398..375R}):
during the first cycle, a reduced number of nodes is considered. Once
the solution cannot be improved further, a second cycle is started
from that point with more nodes. Table~\ref{tab:nodes} summarizes the
number of nodes used in each of the cycles.

\begin{table}
\caption{Number of nodes used in each cycle of our inversions for
  temperature, line-of-sight (l.o.s) velocity and longitudinal magnetic
  field ($B_{long}$).
\label{tab:nodes}}
\centering
\begin{tabular}{c | c c c}
\hline\hline
\ & Cycle 1 & Cycle 2 & Cycle 3\\
\cline{2-4}
Temp & 4 & 5 & 5 \\
$v_{l.o.s}$ & 2 & 4 & 4\\
$B_{long}$ & 0 & 0 & 1 \\
%$v_{turb}$ & 0 & 0 & 0 \\
\hline 
\end{tabular} 
\end{table}

In our first tests with the number of nodes, we realized that placing
the nodes equidistantly along the depth-scale is not an optimal
choice. After some trial and error, we found that we obtain much
better fits whenever, at least, there is a node placed at $\log \tau_{500} \sim
-4.5$ both in temperature and in velocity.

We have modified \textsc{Nicole} to compute the response functions (see definitions in Section~\ref{sec:formation}) using 
centered numerical derivates instead of the default derivatives based on one-side perturbations of the model parameters. 
This makes the code roughly $33\%$ to $50\%$ slower, but it greatly improves the convergence of the inversion that now reaches more consistently similar values of $\chi^2$ from different randomizations of the initial model parameters. Additionally, inspired by \citetads{2012A&A...548A...5V}, we have
removed some \emph{inversion noise}\footnote{Noise in the inferred 2D
maps partly from non-smooth convergence of adjacent pixels.} by
inverting the data once, then smoothing horizontally the
parameters of the model at the locations of the nodes, and restarting
the inversion again. This trick only seems to help in pixels where the
inversion clearly got stuck in a local minimum.

\subsection{Synthetic observation from the 3D simulation}\label{sec:sim}
To study the formation of the \ion{Ca}{2}~$\lambda 8542$ line we use
a 3D magnetohydrodynamical simulation performed with the
\textsc{Bifrost} code (\citeads{2011A&A...531A.154G}), which has also been used
in Paper I. Synthetic
full-Stokes $\lambda8542$ observations have been calculated with
\textsc{Nicole}, as described in Sec.~\ref{sec:nic}.

The simulation includes the upper convection-zone, photosphere,
chromosphere and corona, covering a physical domain that extends vertically from
$z_0=-2.6$~Mm to $z_1=14$~Mm from the surface of the
photosphere. Horizontally, the simulation covers approximately
$24\times24$~Mm. This domain is discretized in a grid of 
$504\times 504\times 496$ points, with a horizontal step of $47.6$~km
and a vertical step of $20$~km in the radiative zone ($-400\leq z\leq
5000$~km). In the convection zone and corona the vertical distance
between grid points becomes larger with increasing scale height.

The calculations are performed with periodic boundary conditions in the
horizontal plane, and open boundary conditions at the top and at the
bottom. At the beginning of the simulation, magnetic flux is injected
at the bottom boundary. The injected field is a flux sheet with no twist
oriented along the $y$-axis from $x_0=4$~Mm to $x_1=16$~Mm  with
strength $B=3300$~G. This injection continues for 1 hour 45 minutes
before being turned off. The field rises steadily through the
convection zone to the photosphere before stalling there after about
an hour. Eventually
the field breaks through the photosphere in certain locations and into the chromosphere, forming expanding
bubbles of magnetic field as described in greater detail in paper
I. Later stages of this simulation were used to study small flares by \citetads{2014ApJ...788L...2A}.

We note that compared to the latest runs performed with \textsc{Bifrost},
this simulation does not include the effect of collisions between
neutrals and ions that can influence the temperature stratification in
the chromosphere (\citeads{2012ApJ...747...87K};
\citeads{2012ApJ...753..161M}; \citeads{2015arXiv150302723M}), 
or the effect of hydrogen non-equilibrium
ionization that can increase the amount of free electrons in the
chromosphere, affecting the opacity of chromospheric lines
(\citeads{2007A&A...473..625L}).

\section{Analysis}\label{sec:resu}

\subsection{3D models from non-LTE inversions}\label{sec:invres}
\begin{figure*}[]
  \centering
  \includegraphics[width=\hsize, trim=0 0 0 0cm, clip]{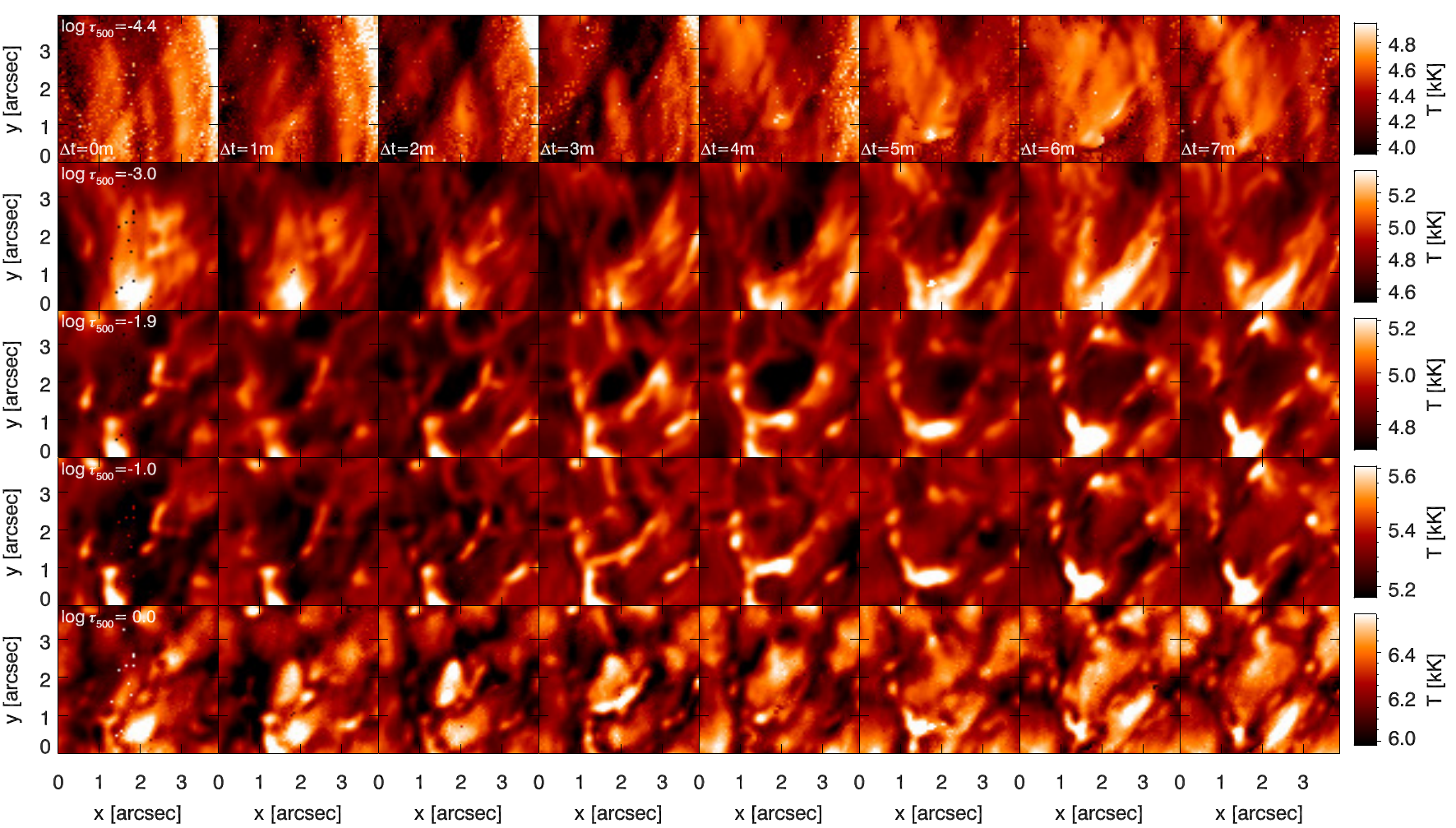}
  \caption{Temporal evolution of the temperature inferred with a
    non-LTE inversion. From left to right, the panels show consecutive
  time steps from our time series. From bottom to top, the panels
  illustrate the inferred temperature at iso-$\log \tau_{500}$
  surfaces in the model. $\Delta t=0$ corresponds to 10:07:16 UT.}
  \label{fig:inv_temp}

  \includegraphics[width=\hsize, trim=0 0 0 3.7cm, clip]{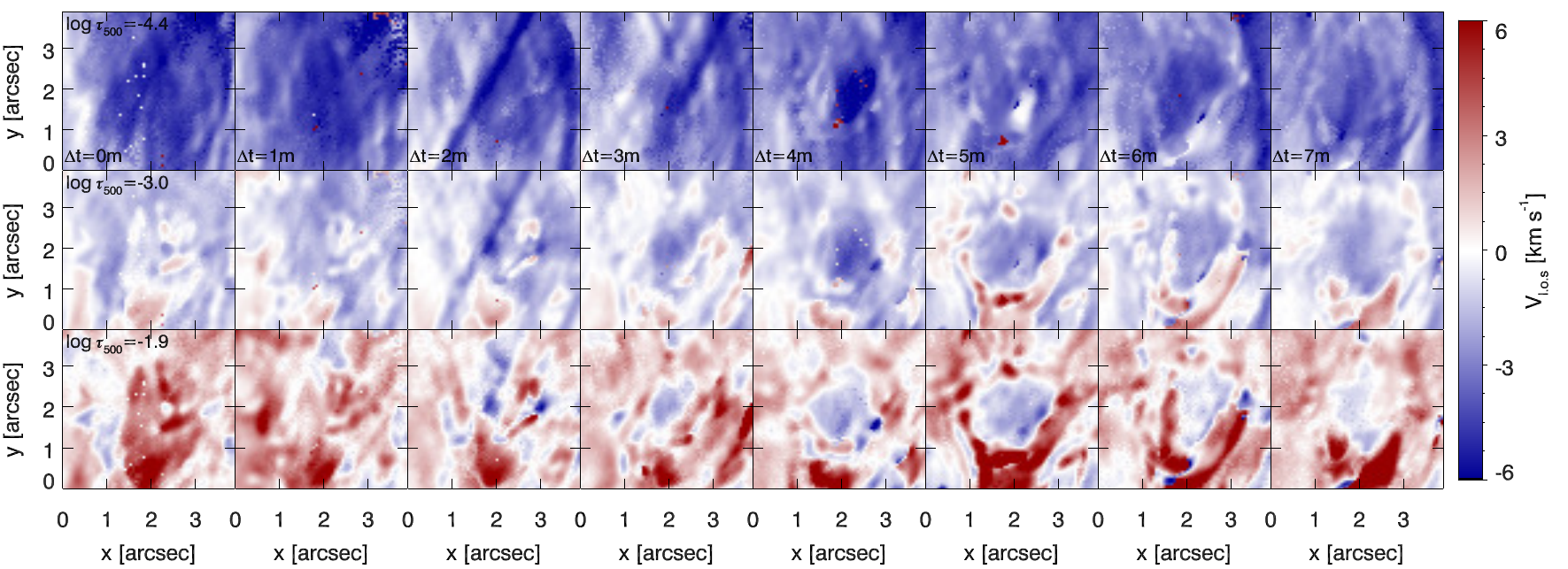}
  \caption{Temporal evolution of the line-of-sight velocity inferred with a
    non-LTE inversion. From left to right, the panels show consecutive
  time steps from our time series. From bottom to top, the panels
  illustrate the inferred temperature at iso-$\log \tau_{500}$
  surfaces in the model. $\Delta t=0$ corresponds to 10:07:16 UT.}
  \label{fig:inv_vel}
  
    \includegraphics[width=\hsize, trim=0 0 0 1.7cm, clip]{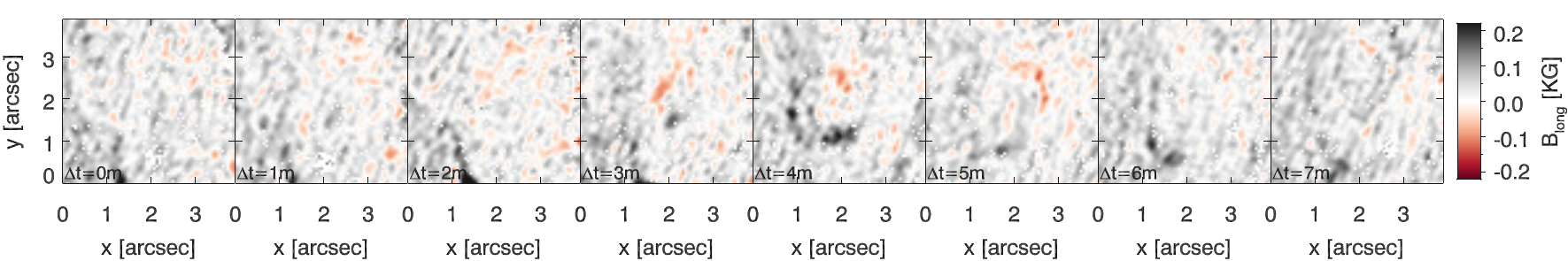}
  \caption{Temporal evolution of the longitudinal component of the magnetic field inferred with a
    non-LTE inversion. From left to right, the panels show consecutive
  time steps from our time series. $\Delta t=0$ corresponds to 10:07:16 UT.}
  \label{fig:inv_blos}
\end{figure*}

In this section we discuss the results of the non-LTE
inversions of the observed \ion{Ca}{2}~$\lambda8542$ Stokes profiles. 
We will focus on the inferred temperature and
line-of-sight velocity maps.

Figure~\ref{fig:inv_temp} illustrates the inferred temperature as a
function of time at $\log \tau_{500} = -4.4, -3.0, -1.9, -1.0,
0.0$. We note that the panels at $\log \tau_{500} = 0.0$ may be an
extrapolation of the model in the upper layers because our
observations do not reach the continuum. Even so, those deep
photospheric layers show typical granulation patterns with hot granules,
colder intergranules and hot bright-points (though geometrically the latter
are probably located in deeper layers than the former, see \citeads{2001A&A...372L..13S}).

At  $\log \tau_{500} = -1.0$ the granulation pattern is reversed
showing hotter intergranular lanes than granules
(\citeads{2007A&A...461.1163C}). 
% viggoh: what ? you have not introduced any loop yet maybe you
% should explain the assumed run of events before starting this description.
Here the footpoints of the bubble are
clearly visible, but otherwise, there are no signatures of the bubble in
the temperature map. Only at $\log \tau_{500} \sim -1.9$, in the center
of the bubble above the granule, does the temperature show a significant
drop compared to the surroundings, especially between $\Delta t=1$~min
and $\Delta t=4$~min. Later on, the temperature rises again in the area where the bubble appeared.

At  $\log \tau_{500} = -3.0$ there is a dramatic change in the
structuring of the temperature, and features that resemble fibrils
start to be visible. Therefore we associate this panel with the lower
chromosphere. 
The magnetic bubble reaches these layers
with a temporal delay of roughly 2 minutes compared to $\log
\tau_{500} = -1.9$, similarly to what was reported in Paper I from the analysis of
images at different line positions. 

Above the lower footpoint (in the image) there
are signatures of heating, quite localized until $\Delta t = 4$~min when
they extend around the bubble. We speculate that this heating can be
produced by three different processes:
\begin{itemize}
\item By magnetic interaction of the footpoints with the \emph{existing} magnetic canopy
of fibrils.
\item By wave dissipation, assuming that waves are channeled through the footpoints and persistently release energy into the chromosphere.
\item By Joule heating at the boundary of the magnetic bubble, where steep gradients in the magnetic field drive electrical currents.
\end{itemize}
Unfortunately, we would need very accurate estimates of the magnetic field at these locations to study whether the brightenings are due to reconnection and/or to estimate the location and effect of electrical currents in our maps. Therefore, this aspect must be clarified in future studies, hopefully with higher sensitivity data.

At $\log \tau_{500} = -4.4$ our inversions show a more homogeneous
temperature structuring, with no signature of the emerging field. Unfortunately,
this event is right at the end of the time-series, so we cannot
confirm whether the bubble manages to reach the middle/upper
chromosphere: in our dataset it does not. The canopy may hinder the ascent of the
bubble and we find evidences of magnetic interaction between the
emerging flux and the existing field especially at the boundary of the
bubble. %(see idealized representation in Figure~\ref{fig:sketch}). 

It is also clear that the apex of the bubble must
be located between $\log \tau_{500} = -3.0$ and
$\log \tau_{500} = -4.4$ in our observations. We note that the panels
in Figure~\ref{fig:inv_temp} may show reduced temperature contrast due
to 3D non-LTE effect (see \citeads{2012A&A...543A..34D}) or to
straylight in the observations. 

Similarly, 2D maps of inferred velocities are presented in
Figure~\ref{fig:inv_vel}. Our results in the deepest layers of the
inverted model seem to be an extrapolation of the velocity field at around  $\log
\tau_{500} \approx -1.5$. We are not surprised because the Lorentzian
(photospheric) wings are likely to be much more sensitive to changes in
temperature than to changes in the velocity field, since the
intensity gradient is not very steep. Close to line
center, the profiles carry more complete information of the velocity
field.

Our inversion results at $\log \tau_{500} = -1.9$ are very similar to those
inferred using \ion{Fe}{1} lines in Paper I. Gas flows up in the
center of the bubble and drains down at the sides. These downflows
coincide with the location of intergranular lanes of photospheric
granulation. At $\log \tau_{500} = -3.0$, the imprint of these
downflows in the surroundings has almost vanished, but in some
locations they are visible at the end of the series, when the bubble has
reached high enough in the atmosphere. Finally, at $\log \tau_{500} =
-4.4$ the images are dominated by the
canopy of fibrils, which appear upflowing along our line-of-sight. The
entire region seems to be moving upwards with speeds of $-6$~km~s$^{-1}$.

We now discuss the inferred longitudinal component of the
magnetic field, shown in Figure~\ref{fig:inv_blos}. The maps at $\Delta t = 0,1,2$~min do not show any feature in pixels harbouring the bubble.
However, from $\Delta t = 3$~min to $\Delta t = 5$~min two patches of opposite polarities appear and slowly separate during the ascent of the bubble.
From $\Delta t = 6$~min, one of the footpoints is still visible at the bottom of the image. The inferred longitudinal field has amplitudes peaking $\pm140$~G which seems to lay just above out sensitivity threshold. This inversion has been performed with a constant magnetic field because the polarization amplitudes are too low to attempt to infer gradients. 

Initially we were surprised to see these opposite polarity patches. 
In Paper I, we used the far wings and the inner core of the 8542 line to estimate the  photospheric and chromospheric fields, respectively. The results show half-moon shaped footpoints in the
photosphere (with opposite polarities), whereas in the chromosphere
we detected signals only in a very small patch above the
footpoints, but unfortunately not in the interior of
the bubble. We suspect this is due to the too restrictive wavelength ranges we used to compute the weak-field approximation both at the core and at the wings, skipping those parts of the line profile where the bulk of the signals are contained.

\begin{figure}[]
  \centering
  \includegraphics[width=\hsize]{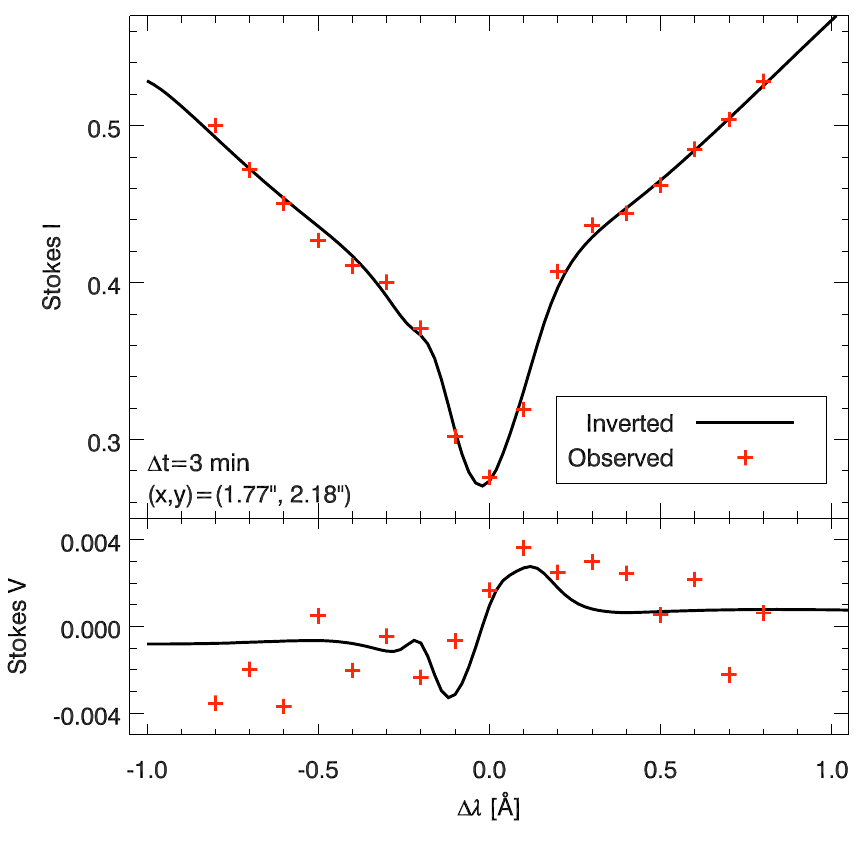}
  \caption{Exemplary observed (red-crosses) and best-fit (black-solid)
    profiles from the center of the
    magnetic bubble at $\Delta t = 3$~min, at the upper opposite polarity patch. Stokes~$I$ (top) and Stokes~$V$ (bottom).}
  \label{fig:fit}
\end{figure}

Now we discuss the quality of the fits to the observed profiles which are reasonably good, but not perfect.
Figure~\ref{fig:fit} shows an example from the upper opposite polarity patch, at $\Delta t = 3$~min. The general aspect of the Stokes~$I$
profile is reproduced by the inversion code, but the sharpest features
in the obsevations appear smoothed in the inversion, an
expected result from the discretization of the nodes in the
inversion (\citeads{2012A&A...543A..34D}): the location and separation
of the nodes are not ideal to catch sharp gradients in
the atmosphere. The Stokes~$V$ profiles are very noisy, but the code seems to find a solution that is compatible with the observations assuming a constant magnetic field. 

We focus now on the strongly tilted blue wing of the profile (see Figure~\ref{fig:time}). We
believe that asymmetry is produced in the observations by the presence
of fibrils. We note that the upper panels in Figure~\ref{fig:inv_vel}
show velocities of the order of $-6$~km~s$^{-1}$ in the fibrils above the
bubble, and velocities are even higher at lower optical-depths. Those velocities correspond to a Doppler shift of
approximately $-180$~m\AA\ from the rest wavelength, which is compatible with the
location of the observed asymmetry in the profile. The asymmetry also
quantifies the difference in velocity between the emerging field and the
fibrils above. We note that in this particular geometry and
heliocentric angle, the fibrils above the bubble appear upflowing, but
a similar event with downflowing fibrils may show a different profile
asymmetry.

\begin{figure*}[]
  \centering
  \includegraphics[width=\hsize]{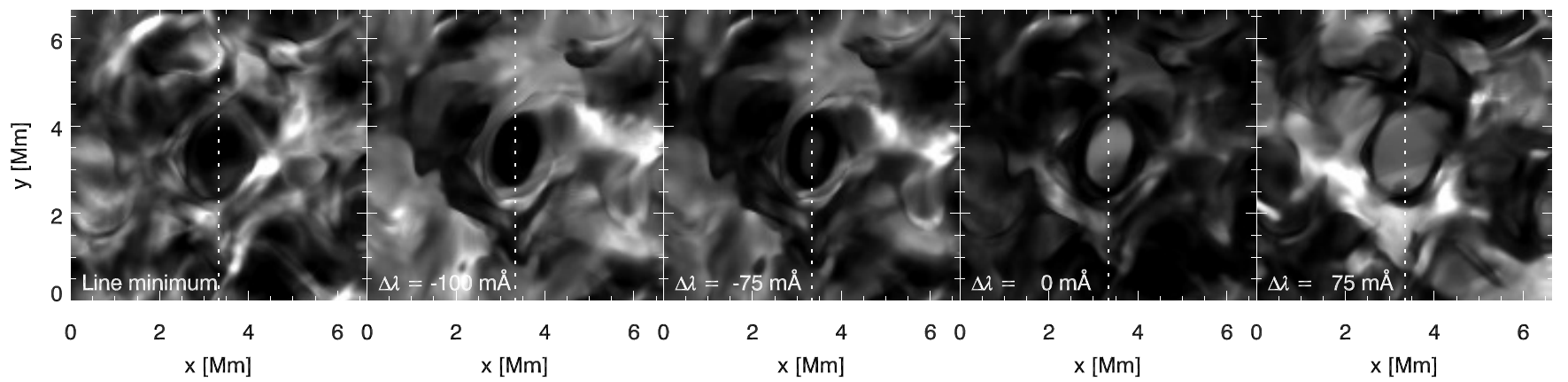}
  \caption{Synthetic intensity images in the $\lambda 8542$ line. The leftmost
    panel represents the line minimum intensity at each pixel. All the
    other panels illustrate, from left to right, monochromatic images
    at $\Delta
    \lambda=-100,-75,0,75$~m\AA, respectively. The vertical dotted line indicates the location of the vertical slices displayed in Figure~\ref{fig:resp}.}
  \label{fig:intsim}
\end{figure*}

\subsection{The formation of the $\lambda 8542$ line within an
  emerging bubble}\label{sec:formation}

The advantage of the MHD simulation is that
we have all the thermodynamical properties in hand, which
makes it possible to study the formation of the line. We have been
puzzled by the peculiar shape of the line profiles inside the
bubble. To find an explanation for such profiles, our first attempt was
to inspect the properties of the 3D simulation.

Figure~\ref{fig:intsim} shows synthetic images in the
\ion{Ca}{2}~$\lambda8542$ line at different wavelengths, for one
time-step of the simulation. In the leftmost panel we illustrate the
line-core intensity for each pixel, and in the consecutive panels,
monochromatic images at $\Delta\lambda=-100,-75,0,75$~m\AA\ relative to
rest wavelength. In the first panel, we have effectively removed
intensity fluctuations due to Doppler motions, and therefore, the
dark circular shape of the flux emerging region becomes clearer. In the
monochromatic panels, the intensity also changes due to Doppler
motions, but it is clear that spectra within the emerging region are
blue-shifted, and therefore, that it is darker at blue
wavelengths. This is similar to what is observed, as described in
Paper I. However, the synthetic profiles are significantly narrower
than those from real observations, with a much steeper Gaussian core
than in the observations. Therefore, the imprint of Doppler shifts is
quite strong in the simulation, and in the red wing there is an intensity enhancement
from Doppler motions that is not obviously visible in the observations.

\begin{figure*}[]
  \centering
  \includegraphics[width=\hsize]{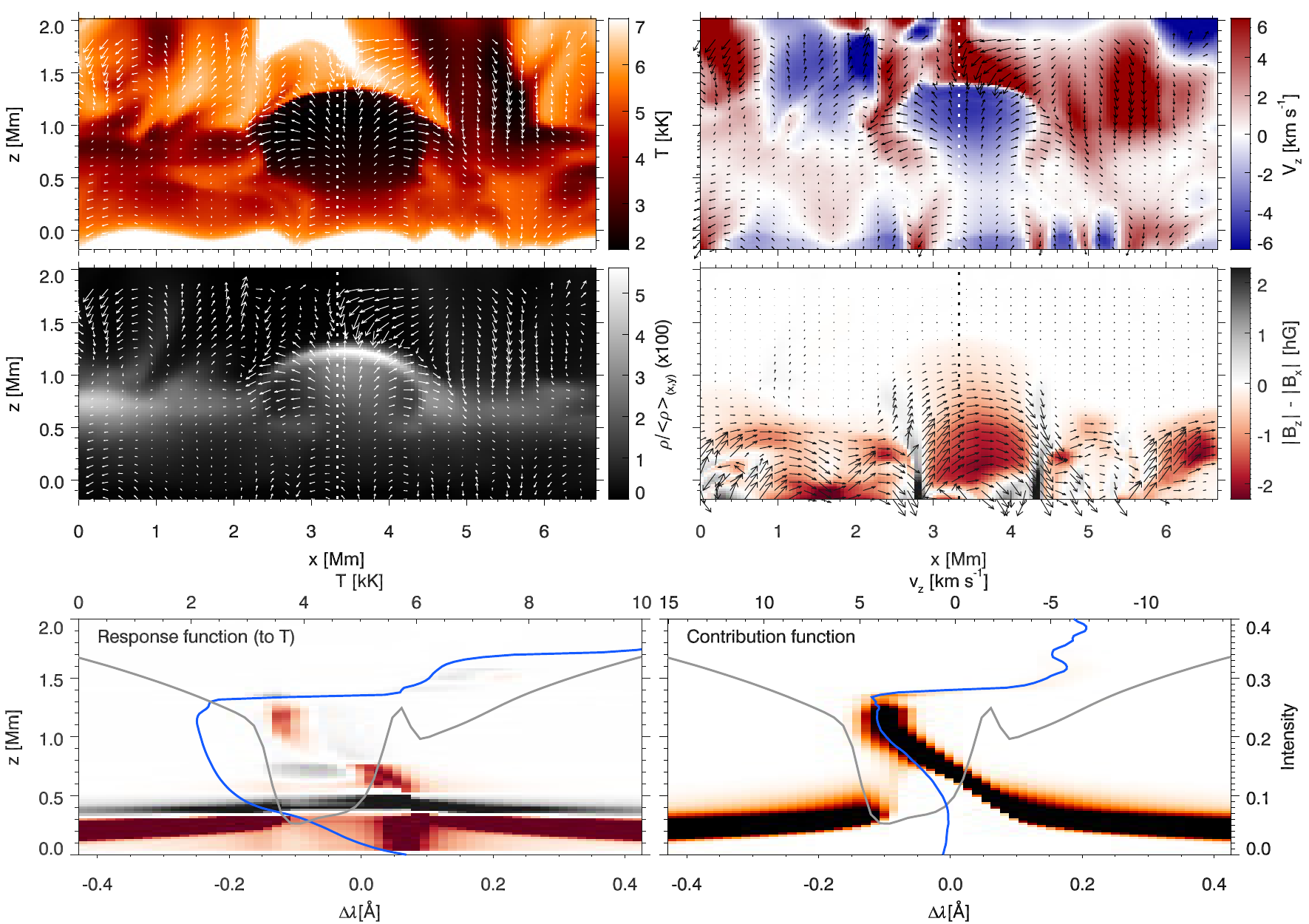}
  \caption{Vertical slice of the temperature (top-left), vertical
    velocity (top-right), density (bottom-left) and magnetic-field
    topology (bottom-right) from a snapshot of the simulation. The
    magnetic field panel illustrates the vertical component in black
    and the horizontal component along the $x$-axis in red.
    The dotted vertical line in the center indicates the location
    where the response and contribution functions have been
    computed. The velocity field has been overplotted with arrows in
    all panels except that showing the magnetic-field strength, where
    the direction of the magnetic field in that plane has been represented instead.
    \emph{Bottom-left:} Response function of the intensity to the
    temperature. \emph{Bottom-right:} Contribution function to the
    intensity. In both cases, the scale has been clipped to enhance
    the upper layers. For clarity, the blue curve in the lowermost panels
    illustrate the corresponding
    temperature (in the leftpanel) and velocity profiles (in the right
    panel) for that column in the 3D model. }
  \label{fig:resp}
\end{figure*}

In the upper and middle rows in Figure~\ref{fig:resp}, we show vertical slices of the corresponding
snapshot from the MHD simulation: in the top row, temperature and
line-of-sight velocity, in the middle row, density\footnote{normalized
to the average density at each horizontal plane.} and
magnetic field strength, indicated in black for vertical fields and in
red for horizontal fields. For completeness, the velocity field has been
over-plotted using arrows in all panels, except in the middle-right
panel, where the arrows indicate the direction of the magnetic field
in the plane of our slices ($Y-Z$). For context, we have indicated the location of these slices in Figure~\ref{fig:intsim} using a dotted white line.

In the simulation, the
emergence of magnetic-flux produces a cold up-flowing
bubble, where the plasma flow is mostly vertical in the center,
and turns almost horizontal on the sides. At the boundary with the
environment, there seems to be a draining effect, and plasma flows
down. We note that density seems to be higher within the emerging region
than in the surroundings, because the gas is compressed at the
upper boundary of the bubble.

To relate the local thermodynamical variables in the simulation to changes in the
intensity profile, we have computed the contribution function to the
intensity and the response function to temperature, to assess whether
the emission feature observed in the Ca II 8542 intensity profile can be related to temperature enhancements at some height in the atmosphere. 

The contribution function ($C_\lambda$) is computed using the monochromatic source
function ($S_\lambda$), and therefore, it encodes information from many
thermodynamical variables that implicitly affect the source
function: 
\begin{equation}
  C_\lambda (z)=S_\lambda(z) \cdot e^{-\tau_\lambda(z)}\cdot \frac{d\tau_\lambda(z)}{dz},
\end{equation}
where $\tau_\lambda(z)$ is the monochromatic optical depth at height
$z$.

The response function ($R_\lambda$) is computed by directly perturbing the physical quantities in
the model, at each layer, and quantifying directly the impact of such a perturbation on the synthetic profile (see e.g.,
\citeads{1977A&A....56..111L}, \citeads{1992ApJ...398..375R},
\citeads{2004ApJ...603L.129S}, \citeads{2007A&A...462.1137O}): 
\begin{equation}
 R_\lambda(z,T)=\frac{\delta I_\lambda(z)}{\delta T(z) \delta z}, \label{eq:resp}
\end{equation}
where $I_\lambda$ is the emerging intensity and $T$ is the temperature at
height $z$ in the model. \citetads{2005ApJ...625..556F} proposed an equivalent method to
compute the response function, which is numerically more stable, with
slightly different form than Eq.~\ref{eq:resp}, that has been used in
our study. We have applied a perturbation of $\delta T(z) = \pm0.01T(z)$.

In the bottom row of Figure~\ref{fig:resp}, we display the $\lambda8542$ line, computed
through the column at the center of the bubble, indicated with
a dashed vertical line in the panels above. The blue curve 
corresponds to the temperature stratification (in the left panel) and
to the velocity stratification (in the right panel). The line profile seems to have
an emission peak in the red wing, similar to what is observed, but
contrary to intuition, the temperature stratification does not exhibit
clear enhancements that could be associated with such peak. Additionally, we
present the response function to the temperature
and the
contribution function for each wavelength
(\citeads{2006ASPC..354..313U}). These two functions quantify how
sensitive the spectral line is to different layers of the model,
however, the contribution function
does not account for radiative coupling of the source function to
non-local properties of the atmosphere.

In our calculations, both methods seem to provide similar results: far
in the wings, the line is mostly influenced by the photosphere, between
0 and 500~km. Close to line center, the line is sensitive to higher
layers. The blue curve represents the temperature stratification with
height (left panel) and the line-of-sight velocity profile
(right-panel). Their corresponding scales are indicated in the upper
axis of each of the panels. Both plots indicate that the core of the
line is formed within the rising magnetic region, which is denser than the
surroundings in the simulation.

The contribution function indicates that the emission peak
in the red wing originates at around 450 km above the surface,
coinciding with the base of the bubble. The velocity profile
also changes there, with a steeper velocity gradient inside the bubble.
The response function shows a
complicated sensitivity to changes in temperature within a large range
of heights, making it very hard to identify a specific regime
responsible for the emission.

\begin{figure}[]
  \centering
  \includegraphics[width=\hsize]{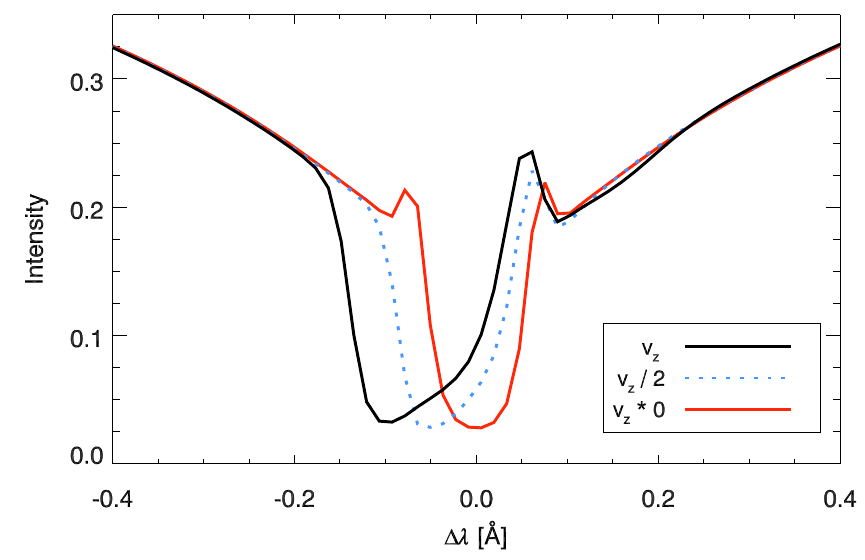}
  \includegraphics[width=\hsize]{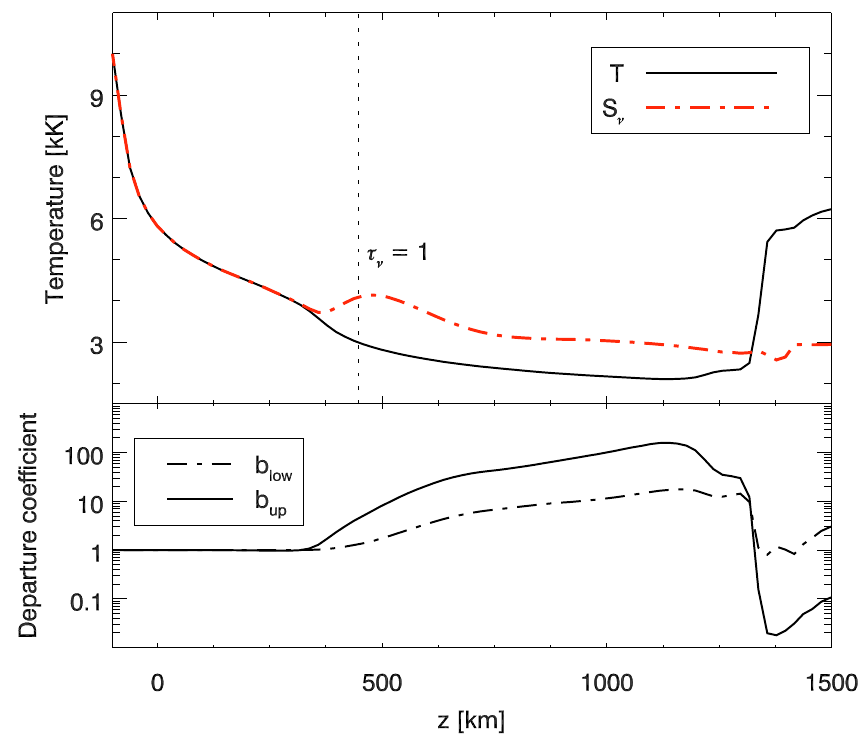}
  \caption{\emph{Top:} Synthetic profiles using the original velocity
     field (solid-black), zero velocities (solid-red) and with the
     velocity field decreased by a factor $\times 2$ (dotted-blue). 
     \emph{Middle:} Temperature profile (solid-black) inside the magnetic bubble
    and the corresponding total source function (red dashed-dotted) at $\Delta \lambda =
    +60$~m\AA. The source function has been converted to radiation
    temperature for comparison with the temperature profile. \emph{Bottom:} Departure coefficients from LTE
      populations for the upper level (solid) and the lower level
      (dashed-dotted) of the transition.}
  \label{fig:source}
\end{figure}

Therefore, we have attacked the problem with a slightly different
approach.
To test whether the emission is an effect an effect of Doppler velocities, we
have run a test where the profile from this column is calculated
using the original velocity stratification, with the velocities set to
zero, and with the velocities decreased by a factor $\times 2$. The
result is illustrated in Figure~\ref{fig:source} (top panel). The red curve is
computed neglecting the velocity field, and therefore it is
symmetric. The emission feature is then visible on both wings of the
line, although there are no temperature enhancements along the
line-of-sight. The plot also demonstrates that the emission continues to be visible when the amplitude of the velocities is half of that in the simulations, but the presence of strong Doppler velocities seem to
enhance the emission in the red wing. The latter is probably because the velocity
gradients (upflowing material) reduce the line opacity in the red wing
(see discussion by \citeads{1984mrt..book..173S}).

To understand the origin of the emission, we have inspected the shape
of the monochromatic source function, and analyzed its dependence on
the local temperature profile. In Figure~\ref{fig:source} (middle panel) we illustrate the temperature profile of
the column and the total source function at $\Delta\lambda =
60$~m\AA, which has been converted to radiation
temperature. The source function decouples from the temperature
profile at around $350$~km from the surface and peaks at
$500$~km. Then it decreases again monotonically. At that wavelength,
the $\tau_\nu = 1$ layer coincides with the bump in the source
function and therefore the emission feature is formed. However, this
behaviour is not due to an increase in the temperature (unlike \citeads{2013ApJ...764L..11D}) or in
the density profiles, rather the opposite. In the lower panel in Figure~\ref{fig:source}, we
show the departure coefficients for the upper and lower levels
of the $8542$ line as a function of height. At $z=500$~km, the upper level departs
from LTE more vigorously than the lower
level, explaining the peak of the source function at that height.  

\citetads{1989A&A...213..360U} reports that such a behaviour a can
occur because the \ion{Ca}{2}~$K$ and the $8542$ lines share the
upper level while their lower levels are collisionally-coupled. Therefore,
the source function of the $8542$ line is set by the $K$ line and
displays the same temperature sensitivity as $3933$~\AA. We emphasize that the
source function in Figure~\ref{fig:source} decouples from the
temperature profile exactly at the temperature drop produced by the magnetic bubble. Given the behaviour observed in the
departure coefficients and the shape of the source function, we have
no reason to think that this is not the same phenomenon.

\begin{figure}[]
  \centering
  \includegraphics[width=\hsize]{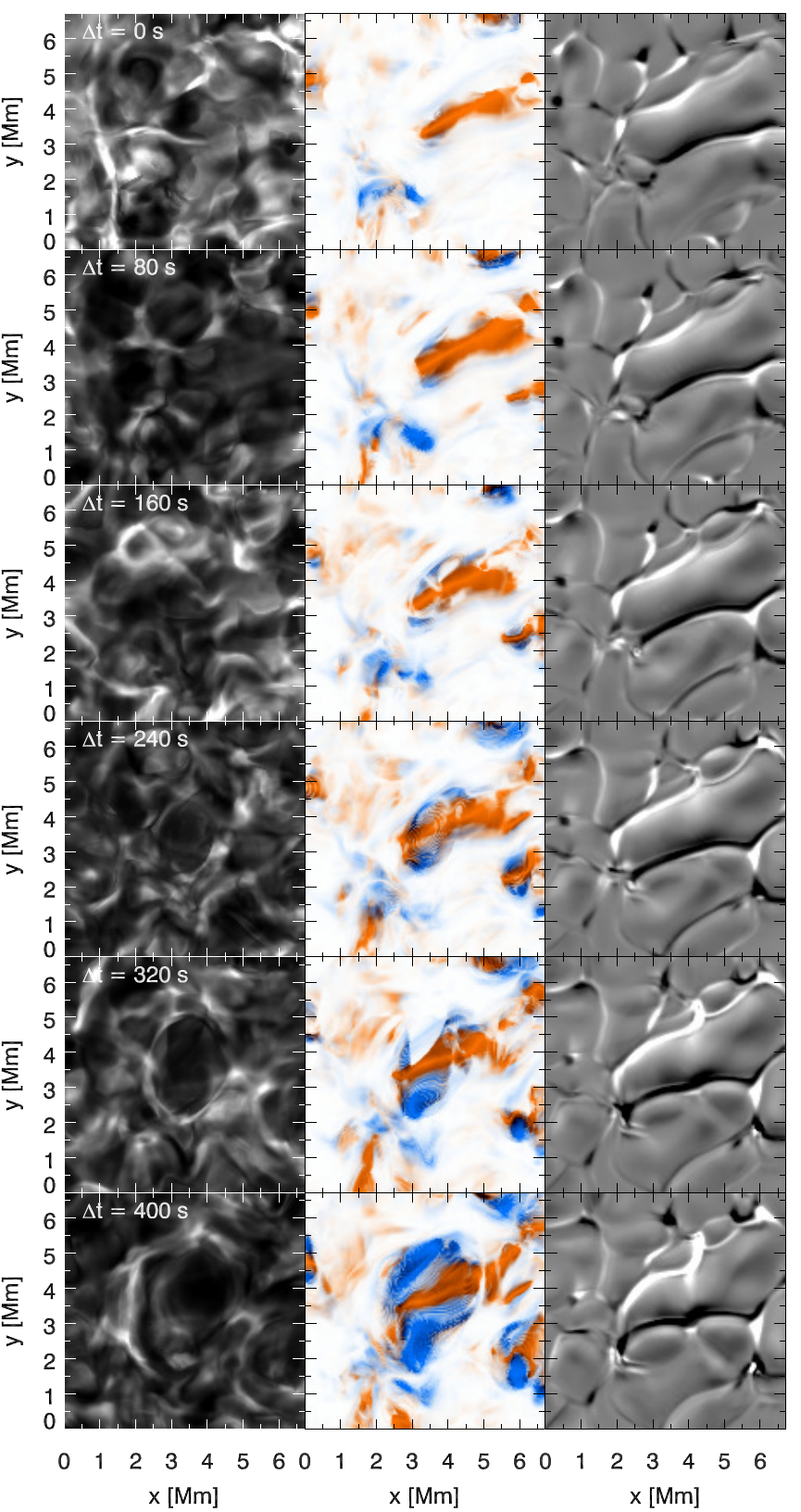}
  \caption{Synthetic observables from our 3D MHD
    simulation. \emph{Left-column}: intensity image constructed with the
    minimum intensity of the line at each pixel. \emph{Middle-column}: color
    composition showing the maximum total linear
    polarization (orange) and the maximum total
    circular polatization (blue). To compose the images, we have
    scaled the total circular and linear
    polarization to the respective maximum values in the selected
    subfield. \emph{Right-column}: monochromatic Stokes~$V$ at
    $\Delta\lambda=-1.83$~\AA\ from line center. Time increases from
    top to bottom.}
  \label{fig:tpol}
\end{figure}

We now focus on the polarization signals in our synthetic
observations. Figure~\ref{fig:tpol} illustrates some non-consecutive
snapshots from the simulation. The left-column displays the minimum
intensity of the profile at each pixel. In the middle-column we have
made a color composite: in orange we represent
$\max(\sqrt{Q_\lambda^2+U_\lambda^2)}$ and in blue $\max(|V_\lambda|)$. In this way, we can
visualize where the field is predominantly inclined. 
% viggoh: some text missing here, I made a guess...
The column on the right illustrates 
Stokes~$V$ in the photospheric wing, and provides information about
the footpoints of our emerging region. In the beginning of the time-series, linear polarization
is dominant in the picture, suggesting that the $\lambda 8542$ line is sampling the
top of the bubble where the field is mostly inclined. As time goes
on and the bubble reaches higher layers in the atmosphere, the imprint
of vertical fields becomes stronger on the sides, where the magnetic
field connects with the foot-points. This is also consistent with the
results from LTE inversions (using the \ion{Fe}{1}~$\lambda
6301/6302$ lines) presented in Paper I. Since we are
tracking the maximum signal of the Stokes profiles, which can
change wavelength due to Doppler motions or to a change in the
formation regime, the observables depicted in Figure~\ref{fig:tpol} may be sensitive to slightly different heights in the model. 

In our observations we did not detect linear polarization
within the magnetic bubble and circular polarisation levels were just above the detection limit, probably because the magnetic field is
highly inclined, and the sensitivity of Stokes~$Q$ and $U$ to the
field strength is much
lower than for Stokes~$V$. Interestingly, we only detected Stokes V signals when the synthetic Stokes I profile shows emission at around $\Delta \lambda = 60$~m\AA\ (e.g. panel 1 of Figure 10 in Paper I). Therefore, we have inspected the
polarization levels in the Stokes profiles computed from the MHD simulation. We note that the longitudinal component of the magnetic field inferred from the observations, ranges between $100-200$~G, and these values are quite similar to those present in the 3D MHD simulation. 

\begin{figure}[]
  \centering
  \includegraphics[width=\hsize]{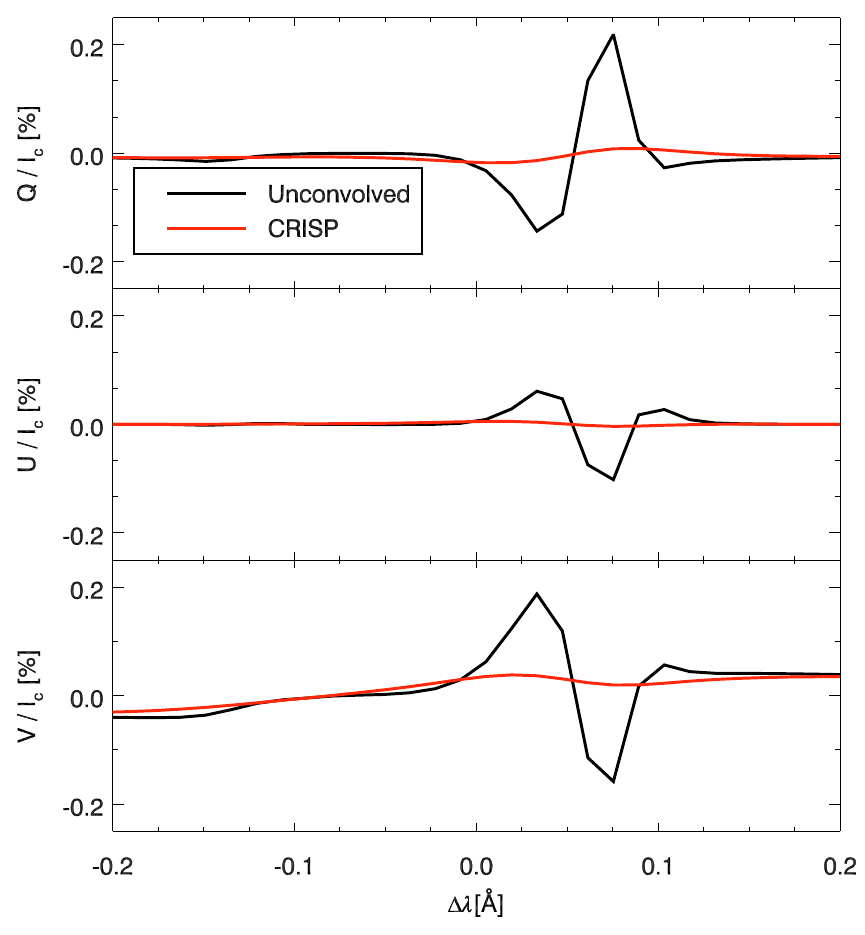}
  \caption{From top to bottom, Stokes~$Q$, $U$ and $V$, convolved with
  a theoretical CRISP transmission profile (red) and unconvolved
  (black). The profiles correspond to the same column from the models
  as it was used in Figure~\ref{fig:resp}.}
  \label{fig:stk}
\end{figure}

The synthetic Stokes~$V$ profiles show polarization levels peaking at
$0.2\%$ relative to the continuum intensity for the linear and
circular components. We have convolved the spectra with a theoretical
CRISP transmission profile at $\lambda8542$ and most of the signal is
washed away, showing peak values one order of magnitude lower than
the unconvolved ones. However, the profiles computed from our simulation are
significantly narrower than the observed profiles (see discussion by \citeads{2009ApJ...694L.128L}, \citeads{2012A&A...543A..34D}, \citeads{2014arXiv1412.1815R}). Therefore, our test with the CRISP transmission
profile is likely to be significantly more pessimistic than reality,
and it should only be taken as a lower limit. Regions with stronger
magnetic-field could produce polarization levels of $\approx
0.1\%$, which is our detection limit. Figure~\ref{fig:stk} illustrates
this experiment. Incidentally, we note that the model shows Stokes~$Q$, $U$ and $V$ signals only at the position of the
emission feature, i.e., the emission brings out polarization signals that would otherwise be
hidden. This is also what happens in our observations.

Summarizing, in this section we have used a 3D MHD simulation to study
the formation of the $\lambda 8542$ within a magnetic bubble emerging into the solar chromosphere. Our
analysis reveals that the emission component in the
red wing of the line is consistent with an increase of the source
function peaking at $z=500$~km. The presence of a strong upflow in the
bubble makes the profile highly asymmetric, hides the emission
feature in the blue-wing and makes the emission in the red-wing stronger.

% \begin{figure}[]
%   \centering
%   \includegraphics[width=\hsize]{fig_tau}
%   \caption{Vertical slices from the 3D MHD simulation in temperature
%     (top), and vertical velocity (bottom). The contours represent iso-$\log
%     \tau_{500}$ surfaces.}
%   \label{fig:tau}
% \end{figure}

\section{Conclusions \& Discussion}\label{sec:conclu}
In this paper we have studied granular-sized flux emergence and the
formation of the $\lambda 8542$ line within an emerging bubble. To this end,
we have performed a multi-pronged analysis using a 3D MHD simulation and
non-LTE inversions of real observations. This study has been driven by
two main goals: to explain the observed $\lambda 8542$ profiles and to
understand the structuring of atmospheric parameters within
flux-emerging regions.

We have performed non-LTE inversions of our observational data,
obtaining 3D estimates of the temperature and the
line-of-sight velocity. Our inferred maps show the temporal evolution
of the temperature stratification as the bubble emerges from the
photosphere and protrudes into the lower chromosphere. The magnetic
bubble leaves a cold imprint in the temperature maps, which is first
visible in the middle/upper photosphere and, with a temporal delay, in
the lower chromosphere. This result is consistent with the analysis
performed in Paper I, using the wings of the 8542 line and LTE inversions of the observed photospheric lines. 
Our inversions also reveal two patches where the magnetic field has opposite polarity. 
These patches separate during the ascent of the bubble, and they seem to trace the more vertical field of the footprints in the upper photosphere, where the bubble is visible. The inferred longitudinal field peaks at $\pm 140$~G.

Our analysis of the 3D MHD simulation has provided valuable
information to understand the peculiarities of the spectral profiles
that are observed within the emerging region. The emission feature that
is present in the red wing of the $\lambda 8542$ line can be explained
by the coupling of the $\lambda 8542$ source function with the
\ion{Ca}{2}~$K$ line, that can pump electrons to the upper level of
the transition. Furthermore, the velocity maps from non-LTE inversions
allow us to relate the asymmetry observed in the blue-wing of the
profile with the canopy of fibrils above the bubble.

In the 3D MHD simulation, the magnetic region emerged without the
presence of an organized large
scale magnetic field in the chromosphere and could therefore reach
higher layers. 
In the observations the bubble seems to get stuck once it
reaches the canopy of fibrils. We believe this is the reason why the
bubble is visible at line center in our synthetic \ion{Ca}{2} 8542 observations, but not in real observations (except perhaps for weak hints of its presence). 

We identify signatures of chromospheric
heating in the lower chromosphere, but we have not been able to identify the heating mechanism that produces these signatures. Also, we have not been able to assess whether the magnetic region continues
rising into the upper chromosphere or if it is destroyed by
interaction with the existing chromospheric magnetic field by
reconnection. These aspects must be addressed in future studies, perhaps
including lines that are sensitive to the upper chromosphere and transition region (like the
\ion{Mg}{2}~$H$ \& $K$ lines, see e.g., \citeads{2013ApJ...772...90L} and the \ion{Si}{4} lines) and to the corona. This will be the subject of Paper III of the present series (Ortiz et al. in prep.), where we will analyze simultaneous observations of flux emergence events with the SST and the IRIS space mission (\citeads{2014SoPh..289.2733D}).

\begin{acknowledgements}
\small
The authors of this paper thank J. Leenaarts and L. Rouppe van der Voort for illuminating discussions. 
J. de la Cruz Rodr\'iguez acknowledges financial support from the
\textsc{Chromobs} project funded by the Knut and Alice Wallenberg
Foundation. L. Bellot Rubio is funded by grants AYA2012-39636-C06-05 and ESP2013-47349-C6-1-R of the Spanish Ministerio de Econom\'ia y Competitividad, including a percentage from European FEDER funds.
The Swedish 1-m Solar Telescope is operated on the island of La Palma
by the Institute for Solar Physics of Stockholm University in the
Spanish Observatorio del Roque de los Muchachos of the Instituto de
Astrof\'isica de Canarias. 
The radiative transfer computations and inversions were
performed on resources provided by the Swedish National Infrastructure
for Computing (SNIC) at the National Supercomputer Centre (Link\"oping
University) and at the High Performance Computing Center North (Ume\aa \ University) with project id \emph{snic2014-1-273}. 
Part of this study was discussed in meetings of the group
\emph{Heating of the Magnetic Chromosphere} at the International Space
Science Institute (ISSI) in Switzerland. 
This research has made use of NASA's
Astrophysics Data System Bibliographic Services.
\normalsize
\end{acknowledgements}

%\bibliography{jaime}

\end{document}